\documentclass[aps,prl,twocolumn,groupedaddress,showpacs,floatfix]{revtex4}
\usepackage{amssymb}

\usepackage{graphicx}

\begin{document}

\title{Joint free energy distribution in the random directed polymer problem}

\author{V.S.\ Dotsenko$^{\, a,d}$, L.B.\ Ioffe$^{\, b}$,
V.B.\ Geshkenbein$^{\, c,d}$, S.E.\ Korshunov$^{\, d}$, 
and G.\ Blatter$^{\, c}$}

\affiliation{$^a$LPTL, Universit\'e Paris VI, 75252 Paris, France}

\affiliation{$^b$Department of Physics and Astronomy, Rutgers University,
   Piscataway, NJ 08854, USA}

\affiliation{$^c$Theoretische Physik, ETH-Zurich, 8093 Zurich,
   Switzerland}

\affiliation{$^d$L.D.\ Landau Institute for Theoretical Physics,
   119334 Moscow, Russia}

\date{\today}

\begin{abstract}
We consider two configurations of a random directed polymer of length $L$
confined to a plane and ending in two points separated by $2u$. Defining the
mean free energy $\bar F$ and the free energy difference $F'$ of the two
configurations, we determine the joint distribution function ${\cal
P}_{L,u}(\bar F,F')$ using the replica approach. We find that for large $L$
and large negative free energies $\bar F$, the joint distribution function
factorizes into longitudinal (${P}_{L,u} (\bar F)$) and transverse
(${P}_{u}(F')$) components, which furthermore coincide with results
obtained prevously via different independent routes.
\end{abstract}

\pacs{
      05.20.-y  %Classical Statistical Mechanics
      75.10.Nr	%Spin-glass and other random models
      74.25.Qt	%Vortex lattices, flux pinning, flux creep
      61.41.+e	%Polymers, elastomers, and plastics
     }

\maketitle

\begin{figure}[b]
   \includegraphics[width=8.0cm]{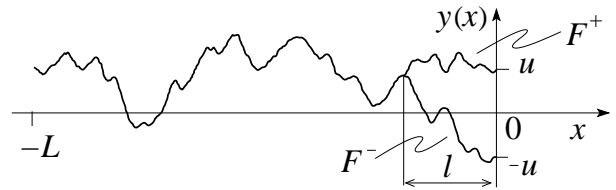}
   \caption[]{Illustration of thermally averaged trajectories 
   $\langle y(x)\rangle_\mathrm{th}$ of a random directed polymer
   in a fixed disorder potential $V(x,y)$. We let the polymer start in 
   an arbitrary position at $x = -L$ and fix the 
   displacement $y=u$ at $x=0$.  Forcing the polymer to end in $y=-u$ 
   produces an alternative average trajectory on a distance $\sim l$.
   Our focus then is on the calculation of the joint distribution 
   function ${\cal P}_{L,u}(\bar F,F')$ for the mean free energy $\bar F
   = (F^++F^-)/2$ and the free energy difference $F'=F^+-F^-$ of the 
   two configurations.}
   \label{fig:setup}
\end{figure}

Directed polymers subject to a random potential exhibit non-trivial
behavior deriving from the interplay between elasticity and disorder; numerous
physical systems can be mapped onto this model and the topic has been the
subject of intense studies \cite{hh_zhang_95}. Despite its undisputable
importance, our knowledge on this generic problem is still limited.
Traditionally, the main focus is on the free energy distribution function, for
which two types of analytical solutions are known for the $(1+1)$-dimensional
case, a polymer confined to a plane (see Fig.\ \ref{fig:setup}): one class
addresses the `longitudinal' problem and determines the distribution function
${\cal P}_L(F)$ for the free energy $F$ of a polymer of length $L$ and fixed
endpoint $y=0$ \cite{kardar_87,zhang_89,spohn_00,monthus_04,kk_06}, while the
other concentrates on the `transverse' problem aiming at the distribution
function ${\cal P}_u(F')$ involving the free energy difference
$F'=F^+\!-\!F^-$ between two configurations with endpoints at $y=\pm u$
\cite{hhf_85,parisi_90,spohn_02}, assuming no dependence on the mean energy
$\bar F\!=\!(F^+\!+\!F^-)/2$ in the limit $L\!\to\!\infty$.  Both approaches have
been helpful in finding the wandering exponent $\zeta$ \cite{hh_85} of
transverse fluctuations $\delta y(L) \propto L^{\zeta}$ of the polymer. On the
other hand, questions how the result for ${\cal P}_u(F')$ is approached from
finite $L$ and how the transverse and longitudinal problems are interrelated
have remained unclear; it is the purpose of this letter to shed light upon
these issues.

Here, we generalize the task of finding the free energy distribution function
for a polymer of length $L$ by studying two configurations of the string
ending in two points separated by $2u$, see Fig.\ \ref{fig:setup}, and
treating both the {\it mean} free energy $\bar F$ and the free energy {\it
difference} $F'$ as relevant variables. The two-point object $F'$ relates to
the natural variable appearing in the Burgers problem \cite{hhf_85}, while for
$u=0$ the variable $\bar F$ reduces to the free energy $F$ of a single
configuration studied in Refs.\
\cite{kardar_87,zhang_89,spohn_00,monthus_04,kk_06}. Our new scheme then
should allow us to place the previous results for ${\cal P}_L(F)$ and ${\cal
P}_u(F')$ into a common context.  Using the replica approach, we determine the
{\it joint} distribution function ${\cal P}_{L,u}(\bar F,F')$ and prove (for a
$\delta$-correlated disorder potential) the separation ${\cal P}_{L,u}(\bar
F,F')={P}_{L,u} (\bar F) \, {P}_{u}(F')$ in the limit of large $L$ and for
large negative values of the mean free energy $\bar F$.  Furthermore, we
derive the form of the two factors ${P}_{L,u}(\bar F)$ and ${P}_u(F')$: on the
one hand, we find that ${P}_{L,u}(\bar F)$ has the same form as Zhang's tail
\cite{zhang_89} for ${\cal P}_L(F)$. On the other hand, to our surprise, we
find that the transverse part ${P}_u(F')$ exactly coincides with the
stationary distribution function ${\cal P}_u(F')$ of the Burgers problem
\cite{hhf_85}, although our solution is associated with rare events in the
far-left tail, while the result of Ref.\ \cite{hhf_85} describes an
equilibrium situation reached in the limit $L\to\infty$. In the following, we
first describe the previous replica analysis leading to the distribution
function ${\cal P}_L(F)$ and its potential pitfalls and then proceed with the
derivation of the joint distribution function ${\cal P}_{L,u}(\bar F,F')$.

We consider an elastic string (elasticity $c$) directed along 
the $x$-axis within an interval $[-L,0]$ and subject to a disorder
potential $V[x,y]$ driving the displacement field $y(x)$,
see Fig.\ \ref{fig:setup}; its energy is given by 
\begin{equation}
   \label{Ham}
   H[y(x);V] = \int_{-L}^{0} dx
   \Bigl\{\frac{c}{2} \bigl[\partial_x y(x)\bigr]^2
   + V[x,y(x)]\Bigr\}.
\end{equation}
The disorder average is carried out over a Gaussian distribution with zero
mean $\langle V(x,y)\rangle_{\scriptscriptstyle V}=0$ and a
$\delta$-correlator $\langle{V(x,y)V(x',y')}\rangle_{\scriptscriptstyle V}
= U_0\, \delta(x-x') \delta(y-y')$.

The standard procedure \cite{kardar_87} leading to the distribution function
${\cal P}_L (F)$ starts from the partition function (we set the Boltzmann
constant $k_B$ equal to unity)
\begin{equation}
\label{dp7}
   Z(L;V) = \int_{y(-L)=0}^{y(0)=0}
   \!\!\!\!\!\!\!\! {\cal D} [y(x)] \, \exp(-H[y(x);V]/T),
\end{equation}
providing us with the free energy $F(L;V) = -T {\rm ln} Z$. The $N$-fold
replication of the polymer and subsequent averaging over disorder realizations
$V$ maps the problem to $N$ quantum bosons with local interactions $-U_0
\delta(y-y')/T$.  In the large $L$ limit, the replica partition function
$Z_r(N;L) = \langle Z^N (L;V) \rangle_{\scriptscriptstyle V}$ is dominated
by the ground state of the quantum problem, which can be obtained from a Bethe
Ansatz solution \cite{McGuire_64}, $Z_r(N;L) \propto \exp[-(N\langle F
\rangle_{\scriptscriptstyle V}+e_3N^3 L)/T]$ (with $e_3=-cU_0^2/ 24T^4$).
Exploiting the relation between the replica partition function $Z_r(N;L)$ and
${\cal P}_L(F)$ as given by the bilateral Laplace transform
\begin{equation}
   \label{fL}
   Z_r(N;L) = 
   \int_{-\infty}^{\infty} \!\!\!\!\! dF \,
   {\cal P}_L (F) \, \exp(-N F/T)
\end{equation}
allows one to show \cite{zhang_89} that the far-left tail of the distribution
function ${\cal P}_L(F)$ assumes the form ${\cal P}_L(F) \propto \exp[-(2/3)(|F|
/F_\ast)^{3/2}]$ with the characteristic free energy scale $F_\ast =(c U_0^2
L/8T^2)^{1/3}\propto L^{1/3}$.  

In Ref.\ \onlinecite{kardar_87}, an attempt has been made to use the result
for $Z_r(N;L)$ and extract the third moment of the distribution function
${\cal P}_L(F)$. While predicting a wrong pre\-factor \cite{krug_92}, this
approach also misses to produce results for other moments.  The reason for
this failure was identified by Medina and Kardar \cite{Medina_93,Kardar_LH},
who pointed out that the two limits $L\to\infty$ (allowing to ignore excited
states) and $N\to 0$ (providing the irreducible moments $\langle\langle
F^k\rangle\rangle_{\scriptscriptstyle V} = (-T)^k \partial^k_N \ln \langle
Z_r(N;L)|_{N\to 0}$ of the distribution function) do not commute. To obtain
estimates for moments, the assumption has to be made that the distribution
function ${\cal P}_L(F)$ is governed by a unique free energy scale $F_\ast
\propto L^{1/3}$; although this assumption cannot be expected to work for the
very distant (non-equilibrium) tails, it turns out that its validity indeed
extends to the far-left tail in the present problem, but does not for the
far-right tail \cite{kk_06}.  Summarizing, the original Bethe Ansatz solution
\cite{kardar_87} allows one to find the (far-left) tail of the distribution
function \cite{zhang_89} but cannot {\it a priori} provide information on its
body \cite{Medina_93,Kardar_LH,Bouchaud_90} as this requires knowledge of the
behavior of $Z_r(N;L)$ for $N \to 0$.

Here, we study a different setup involving two configurations of a polymer
with length $L$ ending in points separated by $2u$; we define the mean free
energy $\bar F= (F^{\scriptscriptstyle +}+F^{\scriptscriptstyle -})/2$ and
difference $F'=F^{\scriptscriptstyle +}-F^{\scriptscriptstyle -}$, with
$F^{\scriptscriptstyle \pm} \equiv F(L,\pm u;V)$ the free energies of polymers
ending in $y(0) = \pm u$. The quantities $\bar F$ and $F^\prime $ are random
variables and we are aiming for the joint distribution function ${\cal
P}_{L,u}(\bar F,F')$. We define the replica partition function $Z_r'(n,m;L,u)$,
which can be expressed as the bilateral Laplace transform of ${\cal P}_{L,u}
(\bar F,F')$,
\begin{eqnarray}
   &&Z_r'(n,m;L,u)\equiv 
   \langle{Z^{n} (L,u;V) Z^{m}(L,-u;V)}\rangle_{\scriptscriptstyle V}
   \label{Znmpm}\\
   &&
   \; =\Bigl\langle e^{\textstyle{-\frac{nF^{\scriptscriptstyle +}}{T}}}
   e^{\textstyle{-\frac{mF^{\scriptscriptstyle -}}{T}}}
   \Bigr\rangle_{\!{\scriptscriptstyle V}}
   =\Bigl\langle e^{\textstyle{-\frac{(n+m) \bar F}{T}}}
                 e^{\textstyle{-\frac{(n-m)F'}{2T}}}
   \Bigr\rangle_{\!{\scriptscriptstyle V}}\nonumber\\
   \noalign{\vskip -2pt}
   && 
   \; =\!\int_{-\infty}^{+\infty} \!\!\!\!\!\!\!\!\!\!
   d\bar F\, dF^\prime \, {\cal P}_{L,u}(F^\prime\!,\bar F)\,
   e^{-(n+m)\bar F/T}\, e^{-(n-m)F'/2T}.
   \nonumber
\end{eqnarray}
The average over disorder realizations $V$ provides us with the replica
partition function in the form
\begin{eqnarray}
   &&Z_r'(n,m;L,u) 
   =\biggl[\prod_{a=1}^{n}\int^{y_a(0)=u}
   \!\!\!\!\!\!\!\!\!\!\!\!\!\!\!\!\!\!\!\!
   {\cal D} [y_a(x)]\biggr]
   \biggl[\prod_{a=n+1}^{n+m}\int^{y_a(0)=-u}
   \!\!\!\!\!\!\!\!\!\!\!\!\!\!\!\!\!\!\!\!\!\!
   {\cal D} [y_a(x)] \biggr]\nonumber\\
   \noalign{\vskip 2pt}
   &&\qquad\qquad\qquad\qquad
   \times \exp\left(-H_{n+m}[\{y_a (x)\}]/T \right),
   \label{Znm}
\end{eqnarray}
with the replica Hamiltonian
\begin{eqnarray}
   \nonumber
   &&H_{n}[\{y_a(x)\}] =
   \int_{-L}^{0} \!\!\!\! dx \, \Bigl\{\frac{c}{2}
   \sum_{a=1}^{n} \bigl[\partial_x y_a(x)\bigr]^2 \\
   &&\qquad\qquad -\frac{U_0}{2T} \sum_{a,b=1}^{n}
   \delta\bigl[y_a(x)-y_b(x)\bigr] \Bigr\}. 
   \label{Hn}
\end{eqnarray}
The replica partition function $Z_r'(n,m;L,u)$ describes a system with $n+m$
trajectories $y_a(x)$ ($a=1, \dots, n+m$) of which $n$ traces terminate at the
point $u$, while the other $m$ trajectories end in the point $-u$; we adopt
free initial conditions \cite{notefi} at $x=-L$ as implied by the absence of
any restriction on $y_a(-L)$ in (\ref{Znm}). All these trajectories are
coupled by the attractive potential $-U_0\delta(y_a-y_b)/T$ deriving from the
disorder correlator.

We use the standard way \cite{kardar_87} to map the path integral (\ref{Znm})
to a Schr\"odinger problem: allowing the $n+m$ trajectories to end in an
arbitrary point ${\bf y} = (y_1,\dots,y_{n+m})$, we define the wave function
$\Psi({\bf y};x) = Z_r'(n,m;L+x,{\bf y})$ which satisfies the imaginary-time
Schr\"odinger equation $-T \partial_x \Psi({\bf y};x) = \hat{H} \Psi({\bf
y};x)$ with the initial condition $\Psi({\bf y};-L)=1$. The Hamiltonian
reads 
\begin{equation}
   \label{Hamop}
   \hat{H} = -\frac{T^2}{2 c}\sum_{a=1}^{n+m}\partial_{y_a}^2
   - \frac{U_0}{2T}\sum_{a,b=1}^{n+m} \delta(y_a-y_b)
\end{equation}
and describes $n+m$ particles of mass $c/T^2$ interacting via the attractive
two-body potential $-U_0\delta(y-y')/T$. The partition function (\ref{Znm}) is
obtained by a particular choice of the final-point coordinates, $Z_r'(n,m;L,u)
= \Psi({\bf u};0)$ with ${\bf u} \equiv (u_1, \dots, u_n = u;
u_{n+1}, \dots, u_{n+m} = -u)$.

The expansion of $\Psi\bigl({\bf y};x\bigr)$ in terms of eigenfunctions
$\Psi_{K,\alpha} = \exp[iK\sum_a y_a/(n+m)] \psi_\alpha(\{y_a\})$ of
(\ref{Hamop}) involves a center of mass component and a factor
$\psi_\alpha(\{y_a\})$ depending only on relative coordinates $y_a-y_b$.  Our
choice of free initial condition $\Psi({\bf y};-L)=1$ implies a vanishing
center of mass momemtum $K=0$ and our expansion assumes the simplified form
\begin{equation}
   \label{Psiexp}
   \Psi({\bf y};0)=\sum_\alpha c_{\alpha} e^{-E_{\alpha} L/T}
   \psi_{\alpha}({\bf y})
\end{equation}
with $E_\alpha$ the eigenenergies. The coefficients $c_\alpha = \langle
\psi_\alpha|\Psi(-L)\rangle/\langle \psi_\alpha| \psi_\alpha\rangle$ follow
from the initial condition $\Psi({\bf y};-L)=1$ with the scalar product
$\langle\psi|\phi\rangle =\int [\prod_a dy_a]\, \delta\bigl[\sum_a
y_a/(n+m)\bigr] \psi(\{y_a\})\phi(\{y_a\})$.

In the {\it limit of large $L$, fixed $u$, and for integer $n,m\geq 1$} (see
below for a detailed discussion on limits and scaling $u$ versus $L$) the sum in
(\ref{Psiexp}) is dominated by the ground state wave function $\psi_0$, for
which the Bethe Ansatz provides the solution \cite{McGuire_64}
\begin{equation}
   \label{psi_0}
   \psi_0({\bf y}) = \exp\Bigl(-\kappa \sum_{a,b}|y_a-y_b|\Bigr)
\end{equation}
with the inverse length $\kappa = c U_0/4T^3$ and the energy 
\cite{noteE}
\begin{equation}
   \label{E_0}
   E_0(n+m)=- c U_0^2 (n+m)[(n+m)^2-1]/24T^4.
\end{equation}
The normalization $\langle\psi_0|\psi_0\rangle=(n+m) /(4\kappa)^{n+m-1}
\Gamma(n+m)$ and the matrix element $\langle\psi_0|\Psi(-L)\rangle =(n+m)
/(2\kappa)^{n+m-1} \Gamma(n+m)$ provide the result $\Psi({\bf y};0) =
2^{n+m-1}e^{-\beta E_0 L} \psi_0({\bf y})$ and evaluating (\ref{psi_0}) at the
endpoint ${\bf u}$, we obtain the expression $\psi_0({\bf u}) = \exp[-4\kappa
|u| nm]$ and hence
\begin{equation}
   \label{Z'}
   Z_r'(n,m;L,u) = 2^{n+m-1}\, e^{-E_0(n+m) L/T} 
   e^{-4\kappa |u| nm}.
\end{equation}
Rewriting the exponent $4\kappa |u| nm=\kappa|u|[(n+m)^2+(n-m)^2]$, 
we can factorize $Z_r'(n,m;L,u)= Z_r^+(n+m;L,u)$ $Z_r^-(n-m;u)$ with
\begin{eqnarray}
   \label{Z_+f}
   Z_r^+ \!&=&\! 2^{n+m-1}\,e^{-E_0(n+m)L/T}\,e^{-\kappa|u|(n+m)^2}, 
   \\ \label{Z_-f}
   Z_r^- \!&=&\! e^{\kappa |u| (n-m)^2}
\end{eqnarray}
depending only on the variables $n+m$ and $n-m$, cf.\ (\ref{Znmpm}).  Hence we
find that the `transverse' problem described by $Z_r^-(n-m;u)$ can be separated
from the (mainly) `longitudinal' part encoded in $Z_r^+(n+m;L,u)$. This
separation into transverse and longitudinal factors is a central element of
our solution and tells us that the joint distribution function ${\cal P}_{L,u}
(F^\prime\!,\bar F)$ as defined in (\ref{Znmpm}) factorizes as well, ${\cal
P}_{L,u}(\bar{F}, F^\prime)= {P}_{L,u}(\bar F)\, {P}_{u}(F^\prime)$.
Correspondingly, we find that the distribution functions ${P}_{L,u}
(\bar{F})$ and  ${P}_u (F^\prime)$ are related to the factors
$Z_r^+(n+m;L,u)$ and $Z_r^-(n-m;u)$ through the bilateral Laplace transforms
\begin{eqnarray}
   \label{Z_+}
   Z_r^+(n+m;L,u) \!\!&=&\!\! \int_{-\infty}^{+\infty} \!\!\!\!\!\!\!\!\! 
   d\bar{F}\, {P}_{L,u}(\bar{F})\, e^{-(n+m) \bar{F}/T},\\
   \label{Z_-}
   Z_r^-(n-m;u) \!\!&=&\!\! \int_{-\infty}^{+\infty} \!\!\!\!\!\!\!\!\! 
   dF^\prime\, {P}_u(F^\prime )\, e^{-(n-m)F^\prime/2T}.
\end{eqnarray}
We note that the above results could be derived for fixed initial conditions
$y_a(-L) = y^i$ as well, however, in this case the factorization appears
only in the limit $L \to \infty$. Also, the restriction to $m,\,n \geq 1$
limits the accessible values of $\bar F$ to large negative values and
restricts the factorization of ${\cal P}_{L,u} (\bar{F},F')$ to the far-left
tail in $\bar F$.

The expression (\ref{Z_-f}) for $Z_r^-$ has been derived for positive 
integer $n,m \geq 1$ and large $L$; its dependence on $n-m$ defines 
$Z_r^-$ on all integers and simple inspection of (\ref{Z_-}) 
allows us to (uniquely) infer the final expression for the free 
energy distribution function
\begin{eqnarray}
   {P}_{u}(F^\prime )
   = \Bigl(\frac{T}{4\pi c U_0 |u|}\Bigr)^{1/2}
   \exp\Bigl(-\frac{T F'^2}{4 c U_0 |u|} \Bigr).
   \label{PF'}
\end{eqnarray}
Formally, the result (\ref{PF'}) can be obtained via analytic continuation of
$Z_r^-$ into the complex plane and use of the inverse Laplace transform (we
define $\xi_- = (n-m)/2T$)
\begin{equation}
   \label{iL}
   {P}_u (F')=\int_{R-i\infty}^{R+i\infty} \!
   \frac{d\xi_-}{2 \pi i} \, Z_r^-(2T\xi_-;u) \, \exp(\xi_- F'),
\end{equation}
requiring an analytic continuation of $Z_r^-$ to the imaginary axis.  This
procedure leads to the identical result (\ref{PF'}), however, without solid
control on the analytic continuation. The result (\ref{PF'}) coincides with
the Gaussian distribution function for the velocities in the corresponding
Burgers problem \cite{hhf_85}, including all numericals.  This comes as a
surprise and may indicate that the factorization, which we can prove for the
far-left tail, may actually prevail throughout all values of $\bar F$.

Next, we analyze what information on $P_{L,u}(\bar F)$ can be extracted from
$Z_r^+(n+m;L,u)$. For $u=0$, the distribution function $P_{L,u} (\bar F)$
coincides with ${\cal P}_L(F)$, $F = \bar{F}$, and the partition function
$Z_r^+(n+m;L,u)$ with the ground state approximation of $Z_r(N;L)$, $N=n+m$. The
partition function $Z_r^+$ as given by (\ref{Z_+f}) is valid for positive
$N=n+m$ and provides, via (\ref{Z_+}), information on large negative free
energies $F = \bar F$, i.e., the left tail of the distribution function,
${\cal P}_L(F)\propto\exp[-(2/3)(|F|/ F_\ast)^{3/2}]$ as calculated by Zhang
\cite{zhang_89}. Inserting this result back into Eq.\ (\ref{Z_+}) and
evaluating the integral via the method of steepest descent, one finds that the
main contribution to the integral arises from values $F\sim -(F_\ast^3/T^2)
(n+m)^2$; negative free energies such that $-F > F_\ast^3/T^2$ then
correspond to positive values $n+m$ for which we can trust the
expression for $Z_r^+(n+m;L,0)$ and hence for ${\cal P}_L(F)$. Going to finite
$u$, we still can trust our result for $Z_r^+(n+m;L,u)$ provided that $F_{\rm
el} = c u^2/2L \ll |\bar{F}|$ (see below) and we find a factor 
$P_{L,u}(\bar{F})$ of basically the same form as for $u=0$.

In order to assess the regime of validity of our results, we have to study the
contribution to Eq.\ (\ref{Psiexp}) of excited states. For $u=0$, the relevant
excited state is the one with lowest energy; this state is one-fold ionized
\cite{takahashi} and its excitation `energy' is given by $\Delta E\, L/T =
(F_\ast/T)^3 (n+m)(n+m-1)$. With $(n+m)^2 \sim |{\bar F}| T^2/F_\ast^3$,
ground state dominance then requires that ${\bar F} \gg T$ and combining this
condition with the one above we find that $|{\bar F}| \gg \max[F_\ast^3/
T^2,T]$. Introducing the temperature dependent Larkin length $L_c(T) \sim
T^5/c U_0^2$, see Ref.\ \onlinecite{blatter_94}, this condition assumes the
form $|{\bar F}| \gg \max[T,TL/L_c]$.  For large $u$, the most dangerous
excited state involves two free clusters with $n$ and $m$ bound particles;
with an `energy' $E\,L/T=-(F_\ast/T)^3 [n(n^2-1)+m(m^2-1)]/3$ and no tunneling
suppression through the excited state wave function, we find a difference in
exponents $[(F_\ast/T)^3 (n+m)- 4\kappa |u|]nm$, from which we obtain the
condition $|{\bar F}| \gg u^2 c/L$; the combination with the restrictions
obtained before produces the overall condition $|{\bar F}| \gg \max[c u^2 /L,
T, TL/L_c]$. Hence, for $L\gg L_c$, typical excursions $\delta y(L) \propto
L^{2/3}$ are well within the domain of applicability of our results.

In analogy with (\ref{iL}), one might directly apply the inverse Laplace
transform (${\cal L}^{-1}$) to the approximate result $Z_r^+$ as given by
(\ref{Z_+f}).  Dropping terms linear in $n+m$ and choosing $u=0$, one easily 
recognizes the integral representation of the Airy function, ${\cal
L}^{-1}[Z_r^+(n+m)]\propto {\rm Ai}(-F/F_\ast)$. The asymptotics at negative $F=
-|F|$ of the Airy function agrees with Zhang's tail of the distribution
function ${\cal P}_L(F)$, as already noted above. However, pushing the free
energy $F$ to positive values, the characteristic oscillations of the Airy
function are incompatible with the positivity of the distribution function
${\cal P}_L(F)$.

Although the above simplified approach correctly accounts for the center of
mass (COM) degrees of freedom, it still fails to produce a consistent result
for ${\cal P}_L(F)$. This observation is in line with a previous study
\cite{mezard}, where the COM motion was accounted for and a negative mean
square displacement $\langle\langle \delta y^2(L)\rangle_\mathrm{th} \rangle_{
\scriptscriptstyle V}$ was found in the $N \to 0$ limit, but contradicts to
the claim in Ref.\ \cite{Bouchaud_90} that the inclusion of the COM motion
leads to a consistent result. We attribute the severe problems appearing in
the derivation of ${\cal P}_L(F)$ to the impossibility to analytically
continue the ground state approximation of $Z_r(N;L)$ derived for integer
$N>1$ and large $L$ to values $N<1$: at $N=1$, all the spectrum describing the
relative motion between bosons collapses to 0 and the former ground state
energy reappears at $N < 1$ with positive energy, cf.\ (\ref{E_0}). As a
result, there is no control on the relevant excitations in the regime $N < 1$.

While the inconsistencies in the analytical continuation of the replica number
$N = n+m$ across unity are quite prominent in the longitudinal problem of
finding ${\cal P}_L(F)$, they appear much more subtle in the analogous
calculation of the transverse distribution function ${\cal P}_u(F')$:
Following Ref.\ \cite{parisi_90} and setting $n+m=0$ in (\ref{Znmpm}), the
integration over $\bar{F}$ could be trivially done and the inverse Laplace
transform of $Z_r'=Z_r^-/2$, see Eq.\ (\ref{Z_-}), provides a result for ${\cal
P}_u(F')$ which, surprisingly, is correct up to a pre\-factor 1/2.  Although the
missing excitations entail merely a spoiled normalization in this case, the
consequences of dropping excitations are much more drastic when dealing with
fixed initial conditions where the pre\-factor diverges. Since our above
analysis of ${\cal P}_{L,u}(\bar F,F')$ does not rely on the replacement of
$n+m$ by zero it is devoid of these problems.

In conclusion, we have calculated the {\it joint} distribution function ${\cal
P}_{L,u}(\bar F,F')$ for two polymer configurations with endpoints separated
by $2u$, allowing us to discuss the longitudinal and transverse problems on an
equal footing, cf.\ (\ref{Znmpm}). Starting from a modified replica approach,
we make use of the Bethe Ansatz solution of the associated quantum boson
problem: We find separability of the longitudinal and transverse problems at
large lengths $L$, a transverse factor which, to our surprise, coincides with
the stationary distribution in the Burgers problem \cite{hhf_85}, and a
longitudinal factor which agrees with Zhang's tail \cite{zhang_89}. The
validity of these results is limited to large negative values of $\bar F$, a
consequence of keeping only the ground state wave function in the solution of
the quantum problem. For a finite-width random potential correlator these
conclusions remain (approximately) valid at not too low temperatures and not
too large $-{\bar F}$, whereas the decrease in temperature or the increase in
$-{\bar F}$ lead to the disappearance of the factorization in ${\cal
P}_{L,u}(\bar F,F')$.  Further progress, particularly with respect to the
longitudinal problem, seems to rely on a better understanding of the spectral
properties of the quantum-boson problem in the regime $0<n+m<1$.

We acknowledge the hospitality of KITP (VBG) and financial support from the
CTS at ETH Z\"urich and from the NSF Grant No.\ PHY99-07949.

%
%This result agrees with the Gaussian distribution function for the velocities
%in the corresponding Burgers problem \cite{hhf_85}. Its width provides us with
%the typical free energy fluctuations $\delta F^\prime \equiv \bigl(\langle
%{{F^\prime}^2} \rangle_{\scriptscriptstyle V}\bigr)^{1/2} 
%=\sqrt{2U_0 c/T}\,|u|^{1/2}$;
%balancing this disorder energy against the elastic elastic energy $\sim c
%u^2/l$ (with $l(u)$ the distance from $x=0$ to the location where the two
%solutions ending in $\pm u$ start to deviate, see Fig.\ \ref{fig:setup}) we
%can obtain the scaling relation $u(l) \sim (U_0/cT)^{1/3} l^{2/3}$ with the
%wandering exponent $\zeta = 2/3$, including also the proper prefactor.
%


\begin{thebibliography}{99}
\bibitem{noteE} We drop the trivial self-energy term %$-U(0)(n+m)/2T$
   in the result for $E_0$.
\end{thebibliography}
\end{document}